\documentclass[num-refs]{wiley-article}

\usepackage{siunitx}
\usepackage[section]{placeins}
\usepackage{algorithm}
\usepackage{algorithmic}

\papertype{Original Article}
\paperfield{Journal Section}

\title{Drug Supply Chain Optimization for Adaptive Clinical Trials}

\abbrevs{ACT, adaptive clinical trial; PSO, particle swarm optimization.}

\author[1]{Jincheng Pang PhD}
\author[2]{Hong Yan PhD}
\author[3]{Zoe Hua PhD}

\affil[1]{Department of Statistics and Data Science, Washington University in St. Louis, Saint Louis, Missouri, 63130, United States}
\affil[2]{Servier Pharmaceuticals, Boston, Massachusetts, 02210, United States}
\affil[3]{Servier Pharmaceuticals, Boston, Massachusetts, 02210, United States}

\corraddress{Hong Yan PhD, Servier Pharmaceuticals, Boston, Massachusetts, 02210, United States}
\corremail{hong.yan@servier.com}

\runningauthor{Jincheng Pang et al.}

\begin{document}

\begin{frontmatter}
\maketitle

\begin{abstract}
With increasing interest in adaptive clinical trial designs, challenges are present to drug supply chain management which may offset the benefit of adaptive designs. Thus, it is necessary to develop an optimization tool to facilitate the decision making and analysis of drug supply chain planning. The challenges include the uncertainty of maximum drug supply needed, the shifting of supply requirement, and rapid availability of new supply at decision points. In this paper, statistical simulations are designed to optimize the pre-study medication supply strategy and monitor ongoing drug supply using real-time data collected with the progress of study. Particle swarm algorithm is applied when performing optimization, where feature extraction is implemented to reduce dimensionality and save computational cost.

\end{abstract}
\end{frontmatter}

\section{Introduction}
Interactive telephone or web system are commonly used to control drug dispensing and manage site inventories of trial supplies \cite{Ruikar2016-ih}. However, it is difficult to fully estimate medication requirement of sites using traditional and automated methods in the clinical trial recruitment process. In addition, with the increasing interest in adaptive clinical trial designs \cite{Coffey2008}, challenges are present to drug supply chain management which may offset the benefit of adaptive designs. The challenges include the uncertainty of maximum drug supply needed, the shifting of supply requirement, and rapid availability of new supply at decision points. These motivate us to develop models to optimize the decision making and analysis of drug supply chain planning for both traditional and adaptive clinical trials.\\
Burnham et al.(2015)\cite{Burnham2015} pointed out that the operational challenge of drug supply continues to be a barrier preventing greater uptake of adaptive designs. They've discussed several effective strategies for drug manufacturing, labeling, packaging, and randomization are discussed and addressed financial concerns. Chen  \cite{Chen2019} further explored the details of the drug supply chain for adaptive clinical trials and proposed two trial supply chain optimization problems that represent different mindsets in response to trial adaptations. They developed a two-stage stochastic program and mixed-integer nonlinear program to solve the two problems, respectively. However, in the two-stage stochastic program, a set of scenarios representing possibilities in the future, including patient's enrollment, drop-out, change of target sample size and drug consumption, has to be fixed and given before the optimization. And in the mixed-integer nonlinear program, patient's enrollment and responses after treatment are considered as deterministic parameters to prevent from an overly complicated model. Besides, drug shortage and shutdown of clinical sites have a great impact on the supply chain management, which haven't been taken into account. Therefore, inspired by \cite{Chen2019}, we try to develop a model to overcome those limitations and make the program simpler to be applied in real cases.\\
On the one hand, the major goal in supply chain management is to overcome the challenges mentioned above to prevent drug shortage as much as possible, which may leads to the shutdown of the trials; On the other hand, we consider to optimize the total cost in the trial, including production cost, recruitment cost, shipment cost, holding cost, disposal cost, along with any other unexpected cost. Thus, the proposed model contains an optimization tool to balance the trade-off. Moreover, simulations are designed to optimize the pre-study supply strategy and monitor ongoing drug supply using real-time data collected with the progress of study. Particle swarm algorithm is applied when performing optimization, where feature extraction is implemented to reduce dimensionality and save computational cost.\\
In this paper, a sequential supply chain optimization algorithm based on ACT is developed, which can be applied to both trial planning stage and monitoring stage after the trial begins. Prior to the trial, the model will provide a vector of recommended production amount for every treatment. After the trial begins, the model can be implemented in every time point, and it will automatically check if the current time point is designed to resupply or perform optimization or simply update observations (like inventory level). If the current time point is pre-specified to resupply sites, the model will make the decision if resupply is needed and if so, it will calculate resupply amount. If we’d like to re-optimize resupply threshold, the model will find the optimal choice of resupply trigger level and recommended inventory level based on newly updated observations.\\
The rest of the paper is organized as follows. We introduce the supply chain management in adaptive clinical trials in Section 2, including general strategy and process, challenges and difficulties in managing drug supply, and resupply scheme. In Section 3, we propose our methodology based on a two-stage optimization problem, discuss the corresponding update rules and constraints, and summarize the model as an algorithm. Several simulation studies and sensitivity analyses are carried out in Section 4 to evaluate
the performance of the proposed design. We conclude with a discussion in Section 5.\\
\begin{figure}[htbp]
\centering
\includegraphics[width=10cm]{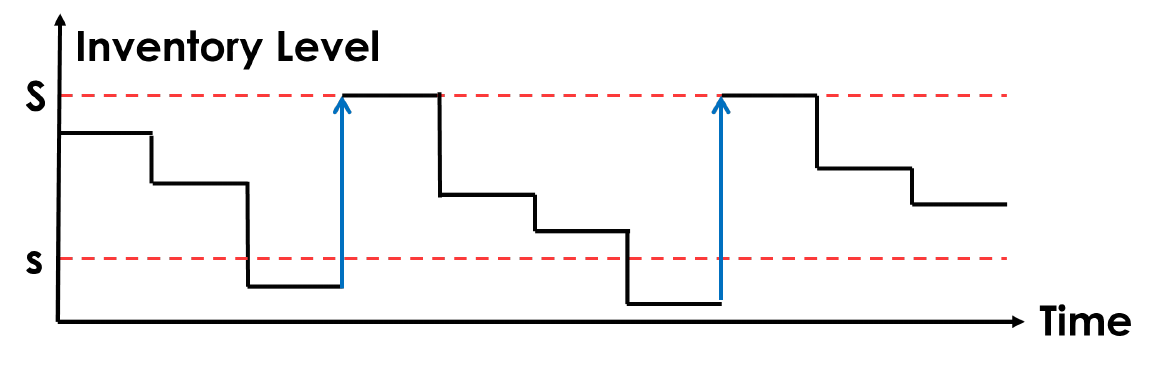}
\caption{This figure illustrates a typical track of inventory level at clinical sites under floor-ceiling resupply scheme. $S$ is the recommended inventory level and $s$ is the resupply trigger level. }
\label{resupplyscheme}
\end{figure}

\section{Supply Chain Management in Adaptive Clinical Trials}
\subsection{Challenges in Supply Chain Management}
Supply Chain management requires careful planning, coordination, and implementation of various strategies to ensure a steady and reliable supply of medication. Prior to the beginning of the trial, accurate demand forecasting is crucial to plan for production schedules, procurement, and inventory management. Establishment of efficient supply chain involves working closely with suppliers, manufacturers, distributors, and logistics partners to ensure timely and reliable delivery of drug. After the trial starts, proper inventory management is critical to prevent shortages or excess stock, where  utilizing inventory management systems and performing regularly monitoring of inventory levels can help streamline operations. Moreover, risk management is applied to identify and mitigate risks to maintain a stable drug supply.\cite{SHAH2004929} During the trial, maintaining compliance with regulations and guidelines helps avoid production delays, recalls, or regulatory actions that could impact drug supply. Besides, continually assessing supply processes and performance helps identify areas for improvement.  Regularly reviewing and optimizing supply chain operations, adopting new practices, and incorporating feedback from colleagues/partners/stakeholders can enhance drug supply management effectively. However, There are several unexpected scenarios that may arise which will impact supply chain management. To be specific, 
\begin{itemize}
    \item Changes in patients enrollment: unforeseen changes in patients enrollment can occurs, such as higher than expected dropout rate or difficulty in recruiting patients. These changes can disrupt the demand forecast for the drug.
    \item Sample size re-estimation: The target sample size can increase during the adaptive trial.
    \item Adverse events or safety concerns: these events can trigger changes in the protocol, requiring adjustments in supply chain to accommodate new safety measures or dosage modifications. Recalls may be necessary which results in potential delays or disruptions in the progress of the study.
    \item Manufacturing or quality control issues: equipment malfunctions or contamination can occur.  These problems may affect the availability or quality of the drug leading to challenges and delays.
    \item Regulatory changes or delays: regulatory authorities play a crucial role in approving trials/licenses. Unanticipated changes in regulatory requirements or delays in obtaining approvals can have significant implications for the supply chain. These changes may require modifications to packaging, labelling, or other aspects of the drug impacting manufacturing or distribution timelines.
    \item Site specific challenges: logistical difficulties, shipping delays, or unexpected storage conditions, can disrupt supply. Need to ensure timely delivery and appropriate storage of drug at each site to maintain integrity.
    \item Global events or emergencies: natural disasters, political unrest, or public health emergencies (COVID-19 Pandemic), can significantly impact supply chain management. These events may disrupt transportation networks, lead to shortages, or cause restrictions on international shipments, making it challenging to maintain a steady supply of drug.
\end{itemize}
\begin{table}[htbp]
\centering
\caption{This table includes basic sets of treatments, clinical sites, and important time points in the model.}
\label{tabset}
\begin{threeparttable}
\begin{tabular}{cl}
\headrow
\textbf{Sets and Indices}&\\
$I$ & The set of treatments indices $i$ \\
$S$ & The set of clinical sites indices $s$ \\
$T$ & The set of time periods $t$; $\left|T\right|$ is the longest duration allowed of the trial.  \\
$TI$ & The set of time points $t$ when interim analysis occurs \\
$TR$ & The set of time points $t$ for resupply \\
$TO$ & The set of time points $t$ for optimization \\
\hline  
\end{tabular}

\begin{tablenotes}
\item All the time points are in the units of weeks. 
\end{tablenotes}
\end{threeparttable}
\end{table}

\subsection{Floor-ceiling Resupply Scheme}
We apply floor-ceiling (S-s) resupply scheme as the inventory control policy in our model\cite{Chen2019}. In S-s policy, for each clinical site and dosage, there are two resupply thresholds $s$ and $S$, namely resupply trigger level and recommended inventory level. When the current inventory level drops below the trigger level $s$, the resupply will be triggered and the resupply amount is designated to fill the inventory up to the recommended inventory level $S$. Note that there exist a shipment lead time and the current inventory level will keep dropping before the resupply arrives. In our model, the thresholds will keep being updated during the trial based on the latest observations and predictions. FIGURE \ref{resupplyscheme} illustrates a typical inventory tracks of the floor-ceiling policy.

\section{Model}
\subsection{Method}
Consider a randomized, double-blinded trial with multiple parallel treatments and several interim analysis time points. Random number of patients will be enrolled at different clinical sites and discrete time periods, and no one can be rejected or missed due to any reason. Enrolled patient will keep consuming drugs in the duration of their treatments. However, in every time period, a random proportion of them will drop out from the trial. The trial will be terminated when there is enough number of patients who enrolled and finished their treatments, i.e. the satisfaction of the target total sample size. Note that the target sample size could change after an interim analysis time point.\\
Then, in this trial, the sequential model we developed will keep monitoring the supply chain and clinical trial. To be specific, in every time point, the model will continually update the observations, like inventory level, to evaluate current status of the supply chain, based on which suggestions will be provided to help decision-making. For instance, the model can tell you whether resuppply is appropriate at current time point to minimize total cost via optimization.\\
In this model, there are three special types of time points during the trial when interim analysis, resupply campaign, and optimization happens.
\begin{table}[htbp]
\centering
\caption{This table provides necessary model settings regarding supply chain and clinical trials, which should be specified before the implementation of the model. }
\label{tabmodelsetting}
\begin{threeparttable}
\begin{tabular}{cl}
\headrow
\textbf{Model Settings}&\\
$c_i^p$ & Production cost of dosage $i$; $\$ /$\textit{dose} \\
$c^{r}_{s}$ & Recruitment and enrollment cost of site $s$; $\$ /$\textit{week}\\
$c^{S}_{s}$ & Shipping cost of site $s$; $\$ /$\textit{box} \\
$c^{h}$ & Holding cost at distribution center; $\$ /$\textit{dose}$/$\textit{week} \\
$c^{h}_{s}$ & Holding cost at site $s$; $\$ /$\textit{dose}$/$\textit{week} \\
$c^{W}_{si}$ & Disposal/recycle cost of dosage $i$ left at site $s$ at the end of trial; $\$ /$\textit{dose}\\
$L$ & Shipment lead time of all dosages and sites; \textit{week}\\
$\tau$ & Time to finish the treatment; \textit{week}\\
$V$ & Volume of a dose; \textit{in}$^3$\\
$Q_{\text{box}}$ & Capacity of a cold shipping box; \textit{in}$^3$\\
$Q_{s}$ & Capacity limit of site $s$; \textit{in}$^3$\\
$P_{\text{shortage}}$ & Penalty parameter on the occurrence of drug shortage\\
\hline  
\end{tabular}

\begin{tablenotes}
\item (A4) makes sure that $c_s^h>c^h$.
\item Shipment lead time $L$ and time to finish the treatment $\tau$ are both positive integers.
\end{tablenotes}
\end{threeparttable}
\end{table}
\subsection{Assumptions \& Model Settings}
Before we talk about the model details, we shall consider several notations and make the following assumptions about the supply chain. TABLE \ref{tabset} summarizes the basic sets and indices in the model.\\
As for the supply management, the immediate availability of drugs is always enforced. Plus, we have the following assumptions\cite{Chen2019}:
\begin{itemize}
    \item[(A1)] No backlog or cross shipment between sites is allowed.
    \item[(A2)] All dosages are produced in one campaign prior to the beginning of the trial and stored at a centralized distribution center with unlimited capacity.
    \item[(A3)] The distances between the DC and clinical sites vary reflected on shipping costs instead of shipping time.
    \item[(A4)] The unit cost of storing drugs is higher at clinical sites than at the DC.
    \item[(A5)] All dosages are shipped together in boxes with same capacity.
\end{itemize}
(A1) is standard assumption that have been commonly used in the clinical trial design. (A2) is applied to simplify the optimization problem. Otherwise, there might be too many decision variables which will drastically increase computational cost. In future research, (A2) can be loosened to generalize the model. (A3) and (A5) can be achieved based on the service policy of logistical company. As for (A4), in most of situations, it is reasonable since with fixed facility cost, the depot with larger storage capacity will have lower unit storage cost compared with clinical sites. If (A4) cannot be guaranteed, the model will still work, as it turns out.\\
Some basic model settings about supply chain and clinical trials are listed in TABLE \ref{tabmodelsetting}, including production cost, recruitment cost, shipping cost, holding cost, disposal cost, shipment lead time, and capacity limit. Note that all these parameters need to be determined before implementing the model.
\begin{table}[htbp]
\centering
\caption{This table provides necessary model inputs regarding supply chain and clinical trials.}
\label{tabinput}
\begin{threeparttable}
\begin{tabular}{cl}
\headrow
\textbf{Inputs}&\\
$n_{st}$ & Number of enrolled patients at site $s$ at time point $t$ \\
$\alpha_{st}$ & Patient drop-out rate at site $s$ at time point $t$ \\
$D_{t}$ & Target sample size of the trial at time point $t$ \\
$\gamma_{it}$ & Average consumption of dosage $i$ at time point $t$ \\
\hline  
\end{tabular}

\begin{tablenotes}
\item $D_{t}$ only changes after the pre-specified time periods where an interim analysis occurs.
\end{tablenotes}
\end{threeparttable}
\end{table}
\subsection{Inputs \& Outputs}
In this proposed model, we need to input $4$ stochastic sequences, listed in TABLE \ref{tabinput}. Those stochastic parameters are observed sequentially during the trial. For example, if current time point is $t^*$, then the observations prior to $t^*$, namely $\left\{n_{st}\right\}_{t=1}^{t^*}$, $\left\{\alpha_{st}\right\}_{t=1}^{t^*}$, $\left\{D_{t}\right\}_{t=1}^{t^*}$, and $\left\{\gamma_{it}\right\}_{t=1}^{t^*}$, will be the inputs in the model at time point $t^*$. TABLE \ref{taboutput} summarizes the outputs of the model. $x_{i}$, $s_{si}$, and $S_{si}$ are three groups of primary decision variables. The rest of them can be viewed as secondary decision variables, which depend on the value of primary variables. In other words, once the primary decision variables are decided, others can be calculated by some update rules (which will be discussed in Section 3.6).\\
Before the trial starts, recommended production amount $x_{i}$ are here-and-now decisions and determined without the realization of future scenarios. After the trial begins, the resupply thresholds (including recommended inventory level $S_{si}$ \& trigger level $s_{si}$) are wait-and-see decisions, and will keep being updated during the trial.

\begin{table}[htbp]
\centering
\caption{This table lists model outputs, including primary and secondary decision variables. }
\label{taboutput}
\begin{threeparttable}
\begin{tabular}{cl}
\headrow
\textbf{Outputs}&\\
\color{red}{$x_{i}$} & Recommended produced amount of dosage $i$; \textit{dose}\\
\color{red}{$s_{si}$} & Trigger level for resupply of dosage $i$ at site $s$; \textit{dose}\\
\color{red}{$S_{si}$} & Recommended inventory level of dosage $i$ at site $s$; \textit{dose}\\
$N_{st}$ & Cumulative number of samples collected from site $s$ at time $t$\\
$y_{it}$ & Inventory of dosage $i$ at distribution center at time $t$; \textit{dose}\\
$y_{sit}$ & Inventory of dosage $i$ at site $s$ at time $t$; \textit{dose} \\
$u_{si}$ & Amount of dosage $i$ shipped to site $s$ before the trial starts; \textit{dose}\\
$u_{sit}$ & Amount of dosage $i$ shipped to site $s$ at time $t$; \textit{dose}\\
$v_{s}$ & Number of boxes used for the shipment to site $s$ before the trial starts\\
$v_{st}$ & Number of boxes used for the shipment to site $s$ at time $t$ \\
$\delta_{t}$ & Binary status of open enrollment; $0$ means terminated\\
$\theta_{t}$ & Binary status of drug supply; $0$ means terminated\\

\hline  
\end{tabular}

\begin{tablenotes}
\item When the primary decision variables, $x_i$, $s_{si}$, and $S_{si}$, are determined, the secondary decision variables can be computed.
\end{tablenotes}
\end{threeparttable}
\end{table}

\subsection{Dimension Reduction}
In Section 3.3, we know that every time we implement the model, it needs to find the optimal choice of three groups of parameters, $x_{i}$, $s_{si}$, and $S_{si}$. Thus, the number of primary decision variables is $\left|I\right|+2\left|S\right|\left|I\right|$, which is also the dimension of the space where the model searches for the optimal result. Depending on the number of treatments and sites, the dimension of the search space could be extremely large, leading to unacceptable computational complexity. However, with one reasonable assumption about drug consumption, the dimension can be reduced remarkably. Consider the following assumption:
\begin{itemize}
    \item[(A6)] Production amount $x_{i}$ linearly depends on the total consumption of drug; Resupply thresholds, $s_{si}$ and $S_{si}$, linearly depend on the average consumption of drug.
\end{itemize}
Given the condition that (A6) is true, then feature extraction can be performed as follows:
\begin{gather}
\underbrace{
\text{Recommended production amount}}_{x_{i}}=\underbrace{
\text{Adjustment multiplier}}_{x_{\text{mul}}}\times \text{Total consumption of dosage}\ i\\
\underbrace{
\text{Recommended inventory level}}_{S_{si}}=\underbrace{
\text{Adjustment multiplier}}_{S_{\text{mul}}}\times \text{Average consumption of dosage}\ i\ \text{at site} \ s\\
\underbrace{
\text{Trigger level}}_{s_{si}}=\underbrace{
\text{Adjustment multiplier}}_{s_{\text{mul}}}\times \text{Average consumption of dosage}\ i\ \text{at site} \ s
\end{gather}
Then, if the drug consumption can be properly estimated, the number of decision variables, i.e. the dimension of search space will be dropped from $\left|I\right|+2\left|S\right|\left|I\right|$ to $3$, since there are only three adjustment multipliers $x_{\text{mul}}$, $S_{\text{mul}}$, and $s_{\text{mul}}$ need to be found in the model. The computational cost can be saved respectively.

\subsection{Drug Consumption Estimation}
\begin{figure}[htbp]
\centering
\includegraphics[width=10cm]{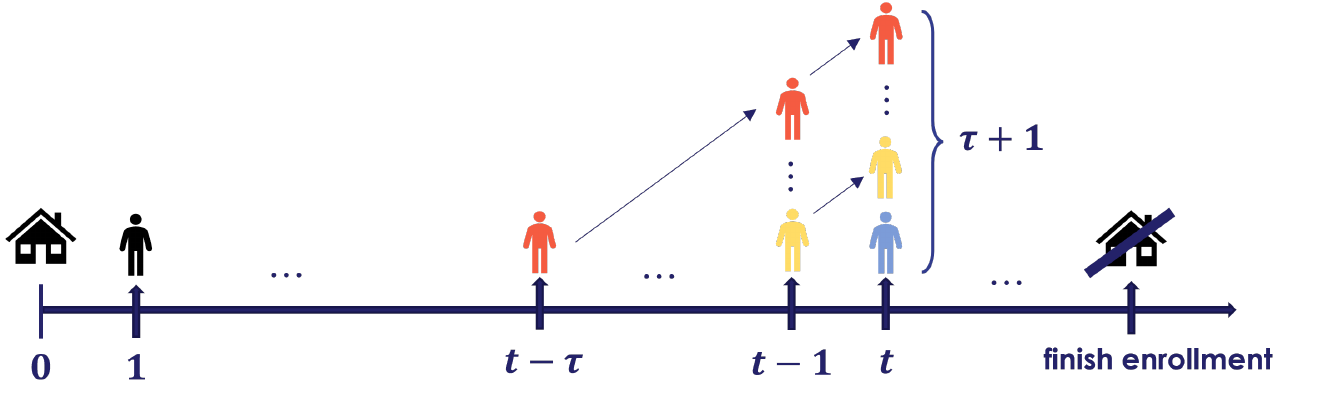}
\caption{This figure shows the composition of patients that consume drugs at time point $t$. Since it takes $\tau$ weeks to finish the treatment, the patients that enrolled at $t-\tau$th week will still consume drugs at week $t$ and there are $\tau+1$ groups of patients consuming drugs at week $t$.}
\label{consumption}
\end{figure}

In order to make sure the dimension reduction applied in Section 3.4 won't cause too much deviation in the model, both total and average consumption of drug during the trial are required to be estimated carefully. Here, we provide the following formula to help calculate the estimation. Define 
$$
d_{sit}:= \text{Amount of dosage}\ i \ \text{consumed in site}\ s\ \text{at time point}\ t.
$$
Then,
\begin{equation}\label{consumptionestimation}
d_{sit} = \begin{cases}\gamma_{i t} \delta_t n_{s t}+\sum_{j=1}^{t-1} \gamma_{i, t-j} \delta_{t-j} n_{s, t-j} \prod_{m=0}^{j-1}\left(1-\alpha_{s, t-m}\right), & \text { if }  1<t\leq \tau \\ \gamma_{i t} \delta_t n_{s t}+\sum_{j=1}^\tau \gamma_{i, t-j} \delta_{t-j} n_{s, t-j} \prod_{m=0}^{j-1}\left(1-\alpha_{s, t-m}\right), & \text { if }  t>\tau \end{cases}
\end{equation}
Note that $d_{sit}$ consists of drug consumption terms contributed by currently enrolled patients and patients who arrived in early periods but have not finished their treatments. Plus, a portion of patients will drop out at every time point, so there exists cumulative multiplication of $1-\alpha_{s t}$. FIGURE \ref{consumption} illustrates where every term in $d_{sit}$ comes from.

\subsection{Optimization}
In this section, we will discuss the details of how the model selects the optimal choice of $x_{i}$, $s_{si}$, and $S_{si}$ in the search space. We aim to minimize the total cost in the future, which can be expressed as follows:
\begin{gather}
    \text{Total cost} = \text{Production cost}+\text{Recruitment cost}+\text{Shipment cost}+\text{Holding cost}+\text{Disposal cost},\ \textit{prior to trial},\\
    \text{Total cost} = \text{Recruitment cost}+\text{Shipment cost}+\text{Holding cost}+\text{Penalty on shortage}+\text{Disposal cost}, \ \textit{after trial begins}.
\end{gather}
Before the trial starts, $x_{i}$, $s_{si}$, and $S_{si}$ are all decision variables. There is no penalty on shortage since we can increase the production amount $x_{i}$ to avoid the occurrence of shortage in the future. After that, the production cost is fixed and unnecessary to be considered in the optimization. However, even with sufficient production, shortage may still happen with certain $s_{si}$ and $S_{si}$, especially in some extreme situations (for example, high enrollment rate). Thus, we add a penalty term in the total cost after trial begins.\\
Prior to the trial, the objective function is
\begin{equation}\label{obj1}
\begin{aligned}
    f_{1}\left(x_i, s_{s i}, S_{s i}\right)&:=\underbrace{\sum_{i \in I} c_i^p x_i}_{\text {Production cost }}+\underbrace{\sum_{t \in T} \sum_{s \in S} c_s^r \delta_t}_{\text {Recruitment cost }}+\underbrace{\sum_{s \in S} c_s^S v_{s}+\sum_{t \in T} \sum_{s \in S} c_s^S v_{s t}}_{\text {Shipment cost }}\\
&+\underbrace{\sum_{t \in T} \sum_{i \in I} c^h \theta_t y_{i t}+\sum_{t \in T} \sum_{s \in S} \sum_{i \in I} c_s^h \theta_t y_{s i t}}_{\text {Holding cost }}+\underbrace{\sum_{s \in S} \sum_{i \in I} c_{s i}^W y_{s i T}}_{\text {Disposal cost }},
\end{aligned}
\end{equation}
where
\begin{gather}\label{feature1.1}
x_i=\underbrace{x_{\text{mul}}}_{\text {Adjustment multiplier }} \underbrace{\sum_{s \in S} \sum_{t \in T} d_{s i t}}_{\text {Total consumption of dosage}\ i},\\
\label{feature1.2}
s_{s i}=\underbrace{s_{\text{mul}}}_{\text {Adjustment multiplier }} \underbrace{\frac{\sum_{t \in T} d_{s i t}}{\sum_{t \in T} \delta_t}}_{\text {Average consumption of dosage}\ i \ \text{at site}\ s},\\
\label{feature1.3}
S_{s i}=\underbrace{S_{\text{mul}}}_{\text {Adjustment multiplier }} \underbrace{\frac{\sum_{t \in T} d_{s i t}}{\sum_{t \in T} \delta_t}}_{\text {Average consumption of dosage}\ i \ \text{at site}\ s},
\end{gather}
based on the dimension reduction method discussed in Section 3.4. Hence, we can denote
$$
\tilde{f}_{1}\left(x_{\text{mul}},s_{\text{mul}},S_{\text{mul}}\right):=f_{1}\left(x_i, s_{s i}, S_{s i}\right)=f_{1}\left(x_{\text{mul}}\sum_{s \in S} \sum_{t \in T} d_{s i t},s_{\text{mul}}\frac{\sum_{t \in T} d_{s i t}}{\sum_{t \in T} \delta_t},S_{\text{mul}}\frac{\sum_{t \in T} d_{s i t}}{\sum_{t \in T} \delta_t}\right).
$$
In the objective function, $\sum_{s \in S} c_s^S v_{s}$ is the shipment cost before trial starts, while $\sum_{t \in T} \sum_{s \in S} c_s^S v_{s t}$ corresponds to the shipment cost during the trial. $\sum_{t \in T} \sum_{i \in I} c^h \theta_t y_{i t}$ and $\sum_{t \in T} \sum_{s \in S} \sum_{i \in I} c_s^h \theta_t y_{s i t}$ represent the holding cost in depot and clinical sites, respectively. $c_{i}^{p}$, $c_{s}^{r}$, $c^{S}_{s}$, $c^{h}$, $c_{s}^{h}$, and $c_{si}^{W}$ are pre-specified supply chain parameters. $\delta_{t}$, $\theta_{t}$, $y_{it}$, and $y_{sit}$ are secondary decision variables that can be calculated sequentially by the update rules (which will be introduced later). $x_{i}$ is primary decision variable. Numbers of boxes shipped, $v_s$ and $v_{st}$, are functions of primary decision variables $s_{si}$, $S_{si}$, secondary decision variable $y_{sit}$, and supply chain parameters $V$, $Q_{\text{box}}$. In fact,
\begin{equation}\label{vs}
    v_{s}=\left\lceil\frac{\sum_{i \in I} V u_{s i}}{Q_{b o x}}\right\rceil, \quad v_{s t}=\left\lceil\frac{\sum_{i \in I} V u_{s i t}}{Q_{b o x}}\right\rceil
\end{equation}
where the amount of dosage $i$ shipped to site $s$ is computed by
\begin{equation}\label{us}
    u_{si}=S_{si},\quad u_{s i t}=\mathbb{I}\left(y_{s i t}<s_{s i}\right)\left(S_{s i}-y_{s i t}\right),
\end{equation}
since according to the floor-ceiling resupply scheme introduced in Section 2.2, $\mathbb{I}\left(y_{s i t}<s_{s i}\right)$ determines whether the resupply is triggered and if so, the resupply amount is $S_{s i}-y_{s i t}$. As mentioned in Section 3.3, secondary decision variables depend on the value of primary decision variables, $x_{i}$, $s_{si}$, and $S_{si}$. Given restrictions (\ref{feature1.1}), (\ref{feature1.2}), and (\ref{feature1.3}), the model aims to find the optimal group of $x_{\text{mul}}$, $s_{\text{mul}}$, and $S_{\text{mul}}$ that minimize the objective function (\ref{obj1}).\\
After trial begins, suppose the trial has been conducted at time point $t^*$. The objective function is
\begin{equation}\label{obj2}
    \begin{aligned}
        f_{2}\left(s_{s i}, S_{s i}\right)&:=\underbrace{\sum_{t>t^*,t \in T} \sum_{s \in S} c_s^r \delta_t}_{\text {Recruitment cost }}+\underbrace{\sum_{t>t^*,t \in T} \sum_{s \in S} c_s^S v_{s t}}_{\text {Shipment cost }}\\
        &+\underbrace{\sum_{t>t^*,t \in T} \sum_{i \in I} c^h \theta_t y_{i t} \mathbb{I}\left(y_{i t}>0\right)+\sum_{t>t^*,t \in T} \sum_{s \in S} \sum_{i \in I} c_s^h \theta_t y_{s i t} \mathbb{I}\left(y_{s i t}>0\right)}_{\text {Holding cost }}\\
        &+\underbrace{\sum_{t>t^*,t \in T} \sum_{s \in S} \sum_{i \in I} P_{\text {shortage }} \frac{\operatorname{sgn}\left(y_{s i t}\right)-1}{2} y_{s i t}}_{\text {Penalty on shortage }}+\underbrace{\sum_{s \in S} \sum_{i \in I} c_{s i}^W y_{s i T} \mathbb{I}\left(y_{s i T}>0\right)}_{\text {Disposal cost }},
    \end{aligned}
\end{equation}
where
\begin{gather}\label{feature2.1}
s_{s i}=\underbrace{s_{\text{mul}}}_{\text {Adjustment multiplier }} \underbrace{\frac{\sum_{t>t^*,t \in T} d_{s i t}}{\sum_{t>t^*,t \in T} \delta_t}}_{\text {Average consumption of dosage}\ i \ \text{at site}\ s},\\
\label{feature2.2}
S_{s i}=\underbrace{S_{\text{mul}}}_{\text {Adjustment multiplier }} \underbrace{\frac{\sum_{t>t^*,t \in T} d_{s i t}}{\sum_{t>t^*,t \in T} \delta_t}}_{\text {Average consumption of dosage}\ i \ \text{at site}\ s}.
\end{gather}
Similarly, denote
$$
\tilde{f}_{2}\left(s_{\text{mul}},S_{\text{mul}}\right):=f_{2}\left(s_{s i}, S_{s i}\right)=f_{2}\left(s_{\text{mul}}\frac{\sum_{t \in T} d_{s i t}}{\sum_{t \in T} \delta_t},S_{\text{mul}}\frac{\sum_{t \in T} d_{s i t}}{\sum_{t \in T} \delta_t}\right).
$$
In the objective function, the holding cost term contains indicator function $\mathbb{I}\left(y_{i t}>0\right)$ since when shortage happens, the inventory level drops below zero with no holding cost. The penalty term is proportional to the shortage level. That is, when inventory level $y_{sit}>0$, $\frac{\operatorname{sgn}\left(y_{s i t}\right)-1}{2}=0$ and there is no penalty term. If $y_{sit}<0$, then $\frac{\operatorname{sgn}\left(y_{s i t}\right)-1}{2}y_{sit}=\left|y_{sit}\right|>0$ is the shortage level. Similarly, numbers of boxes shipped can be calculated as 
\begin{equation}\label{vu}
    v_{s t}=\left\lceil\frac{\sum_{i \in I} V u_{s i t}}{Q_{b o x}}\right\rceil, \quad u_{s i t}=\mathbb{I}\left(y_{s i t}<s_{s i}\right)\left(S_{s i}-y_{s i t}\right).
\end{equation}
After trial begins, there are two groups of primary decision variables, $s_{si}$ and $S_{si}$, which will be reduced to be $s_{\text{mul}}$ and $S_{\text{mul}}$ given restrictions (\ref{feature2.1}) and (\ref{feature2.2}). Thus, the model needs to search for optimal $s_{\text{mul}}$ and $S_{\text{mul}}$ that minimize the objective function (\ref{obj2}).

\subsection{Update Rules \& Constraints}
Suppose the current time point is $t^*$. Based on previous observations of inputs and outputs (including primary and secondary decision variables), namely $\left\{n_{st}\right\}_{t=1}^{t^*}$, $\left\{\alpha_{st}\right\}_{t=1}^{t^*}$, $\left\{D_{t}\right\}_{t=1}^{t^*}$, $\left\{\gamma_{it}\right\}_{t=1}^{t^*}$, $x_{i}$, $s_{si}$, $S_{si}$, and $\left\{N_{st}\right\}_{t=1}^{t^*-1}$, $\left\{y_{it}\right\}_{t=1}^{t^*-1}$, $\left\{y_{sit}\right\}_{t=1}^{t^*-1}$, $u_{si}$, $\left\{u_{sit}\right\}_{t=1}^{t^*-1}$, $v_{s}$, $\left\{v_{st}\right\}_{t=1}^{t^*-1}$, $\left\{\delta_{t}\right\}_{t=1}^{t^*-1}$, $\left\{\theta_{t}\right\}_{t=1}^{t^*-1}$, we could update the value of secondary decision variables $N_{st^*}$, $y_{it^*}$, $y_{sit^*}$, $\delta_{t^*}$, and $\theta_{t^*}$ at current time point $t^*$. The update rules are as follows:
\begin{itemize}
    \item[(R1)] Number of patients off treatment:
\begin{equation}\label{r1}
\begin{array}{cc}
N_{s t}=0 \quad & \forall s\in S,  t \leq \tau,\\
    N_{s t}=N_{s, t-1}+\delta_{t-\tau} n_{s, t-\tau} \prod_{m=0}^{\tau-1}(1-\alpha_{s, t-m}) & \forall s\in S, t>\tau,
\end{array}
\end{equation}
where $\delta_{t-\tau} n_{s, t-\tau}$ is the number of enrolled patients at time point $t-\tau$ and $\alpha_{s, t-m}$ is drop-out rate during $[t-\tau+1, t]$.
\item[(R2)] Enrollment \& supply chain status: 
\begin{equation}\label{r2}
\begin{array}{cc}
\delta_t=1, \theta_t=1 & t=1,\\
    \delta_t=\mathbb{I}\left(\sum_{s \in S} N_{s, t-1}<D_{t-1}\right), \theta_t=\delta_{t-\tau} & \forall t>1,
\end{array}
\end{equation}
where the indicator function means that the enrollment will be closed ($\delta_{t}=0$) when the number of patients who enrolled and finished their treatments is greater than the target sample size. And $\theta_t=\delta_{t-\tau}$ implies that the enrollment status is $\tau$-periods ahead of the supply status.
\item[(R3)] Inventory at distribution center:
\begin{equation}\label{r3}
\begin{array}{cc}
y_{i t}=x_i-\sum_{s \in S} u_{s i}-\sum_{s \in S} u_{s i 1} & \forall i\in I, t=1,\\
    y_{i t}=y_{i, t-1}-\theta_t \sum_{s \in S} u_{s i t} & \forall i\in I, t>1.
\end{array}
\end{equation}
When $t=1$, the inventory level at depot is determined by production amount $x_i$, shipment amount prior to the trial $u_{si}$, and shipment amount at the first time point $u_{si1}$. When $t>1$, current inventory is previous inventory minus shipment amount at current time point.
\item[(R4)] Inventory at clinical sites:
\begin{equation}\label{r4}
\begin{array}{cc}
y_{s i t}=u_{s i}-d_{s i 1} & \forall s\in S, i\in I, t=1,\\
    y_{s i t}=y_{s i, t-1}-d_{s i t} & \forall s\in S, i\in I, t \in(1, L],\\
    y_{s i t}=y_{s i, t-1}+\theta_{t-L} u_{s i, t-L}-d_{s i t} & \forall s\in S, i\in I, t >L.
\end{array}
\end{equation}
When $t=1$, the inventory at sites is computed by subtracting drug consumption $d_{si1}$ from the amount of drug shipped to the site prior to the trial. Note that $L$ is the shipment lead time, which means there is $L$ weeks delay from the depot to clinical sites. Thus, no supply will arrive until $t>L$.
\end{itemize}
Meanwhile, we have some constraints when searching for the optimal decision variables:
\begin{itemize}
    \item[(C1)] Capacity limit at clinical sites: 
\begin{equation}\label{c1}
\begin{array}{cc}
\underbrace{\sum_{i \in I} V u_{s i}}_{\text{shipment prior to trial}} \leq \underbrace{Q_s}_{\text{space limit at site}} & \forall s\in S \\
\underbrace{\sum_{i \in I} V y_{s i t}}_{\text{inventory before first supply}} \leq Q_s & \forall s, t \leq L\\
\underbrace{\sum_{i \in I} V y_{s i, t-1}}_{\text{previous inventory}}+\underbrace{\sum_{i \in I} V u_{s i, t-L}}_{\text{resupply}} \leq Q_s & \forall s, t>L
\end{array}
\end{equation}
    Note that (A2) makes sure that there is no capacity limit for distribution center.
\item[(C2)] Bounds:
\begin{equation}\label{c2}
\begin{array}{cc}
x_i \in \mathbb{R}_{+}  & \forall i\in I \\
\delta_t, \theta_t \in\{0,1\} & \forall t\\
N_{s t}, y_{i t}, y_{s i t}, u_{si}, u_{s i t} \in\left\{0, \mathbb{R}_{+}\right\} & \forall s\in S, i\in I, t\\
v_{s}, v_{s t} \in\left\{0, \mathbb{Z}_{+}\right\} & \forall s\in S, t
\end{array}
\end{equation}
\end{itemize}

\begin{algorithm}[htbp]
	\renewcommand{\algorithmicrequire}{\textbf{Input:}}
	\renewcommand{\algorithmicensure}{\textbf{Output:}}
	\caption{Particle Swarm Optimization in Trial Planning Stage}
	\label{alg1}
	\begin{algorithmic}[1]
            \REQUIRE Stochastic inputs $n_{st}$, $\alpha_{st}$, $D_{t}$, and $\gamma_{it}$; Time point sets $TI$ and $TR$ for interim analysis and resupply; Particle swarm optimization parameters: population size $pop$, dimension $D=3$, maximum number of iterations $i_{\text{max}}$, inertia weight $\omega$, cognitive component $c_1$, social component $c_2$. let $\vec{g}$ be the best known position of the entire swarm. Set $j=0, m=0$.
            \WHILE{$j < pop$}
            \STATE Initialize the particle's position $\vec{x}_{j}=\left(x_{\text{mul}}^{(j)}, s_{\text{mul}}^{(j)}, S_{\text{mul}}^{(j)}\right)$, velocity $\vec{v}_{j}=\left(v_{x}^{(j)},v_{s}^{(j)},v_{S}^{(j)}\right)$ with uniform distribution.
            \STATE Compute $\left\{y_{it}\right\}_{t=1}^{T}$ and $\left\{y_{sit}\right\}_{t=1}^{T}$ from (\ref{r3}), (\ref{r4}).
            \IF{$y_{it},y_{sit}>0$ for all $s,i,t$ (no drug shortage)}
            \STATE Initialize the particle's best known position to its initial position: $\vec{p}_{j}=\vec{x}_{j}$
            \IF{$\tilde{f}_{1}\left(\vec{p}_{j}\right)<\tilde{f}_{1}\left(\vec{g}\right)$}
            \STATE Update the swarm's best known position: $\vec{g}=\vec{p}_{j}$
            \ENDIF
            \STATE j=j+1
            \ENDIF
            \ENDWHILE
            \WHILE{$m < i_{\text{max}}$}
            \FOR{$j$ in $1:pop$}
            \STATE Update the particle's velocity: $\vec{v}_{j}=\omega \vec{v}_{j} + c_{1}U(0,1)\left(\vec{p}_{j}-\vec{x}_{j}\right)+c_{2}U(0,1)\left(\vec{g}-\vec{x}_{j}\right)$
            \STATE Update the particle's position: $\vec{x}_{j}=\vec{x}_{j}+\vec{v}_{j}$
            \IF{$\tilde{f}_{1}\left(\vec{x}_{j}\right)<\tilde{f}_{1}\left(\vec{p}_{j}\right)$}
            \STATE Update the particle's best known position: $\vec{p}_{j}=\vec{x}_{j}$
            \IF{$\tilde{f}_{1}\left(\vec{p}_{j}\right)<\tilde{f}_{1}\left(\vec{g}\right)$}
            \STATE Update the swarm's best known position: $\vec{g}=\vec{p}_{j}$
            \ENDIF
            \ENDIF
            \ENDFOR
            \ENDWHILE
		\ENSURE Optimized decision variables $\vec{g}=\left(x_{\text{mul}}^{\text{best}}, s_{\text{mul}}^{\text{best}}, S_{\text{mul}}^{\text{best}}\right)$.
	\end{algorithmic}  
\end{algorithm}

\subsection{Algorithm}
\subsubsection{Particle Swarm Optimization}
We apply particle swarm optimization (PSO) algorithm \cite{488968} to search for the optimal result. PSO algorithm starts with a population of candidate solutions, called particles, and keep updating the particle's position and velocity in the search space. Each particle's movement is influenced by the particle's own best known position (personal best), and the best known positions in the population (group best), which will be updated as better position is found by other particles.\\
\begin{algorithm}[htbp]
	\renewcommand{\algorithmicrequire}{\textbf{Input:}}
	\renewcommand{\algorithmicensure}{\textbf{Output:}}
	\caption{Particle Swarm Optimization in Trial Monitoring Stage}
	\label{alg2}
	\begin{algorithmic}[1]
            \REQUIRE Current time point $t^*$; Observations $\left\{n_{st}\right\}_{t=1}^{t^*}$, $\left\{\alpha_{st}\right\}_{t=1}^{t^*}$, $\left\{D_{t}\right\}_{t=1}^{t^*}$, $\left\{\gamma_{it}\right\}_{t=1}^{t^*}$, $\left\{N_{st}\right\}_{t=1}^{t^*}$, $\left\{\delta_{t}\right\}_{t=1}^{t^*}, \left\{\theta_{t}\right\}_{t=1}^{t^*+\tau}$, $\left\{d_{sit}\right\}_{t=1}^{t^*}$, $\left\{y_{it}\right\}_{t=1}^{t^*}, \left\{y_{sit}\right\}_{t=1}^{t^*}, u_{si}, \left\{u_{sit}\right\}_{t=1}^{t^*}, v_s, \left\{v_{st}\right\}_{t=1}^{t^*}$; Simulated stochastic inputs $\left\{n_{st}\right\}_{t=t^*+1}^{T}$, $\left\{\alpha_{st}\right\}_{t=t^*+1}^{T}$, $\left\{D_{t}\right\}_{t=t^*+1}^{T}$, $\left\{\gamma_{it}\right\}_{t=t^*+1}^{T}$; Time point sets $TI$ and $TR$ for interim analysis and resupply; Particle swarm optimization parameters: population size $pop$, dimension $D=2$, maximum number of iterations $i_{\text{max}}$, inertia weight $\omega$, cognitive component $c_1$, social component $c_2$. let $\vec{g}$ be the best known position of the entire swarm. Set $j=0, m=0$.
            \WHILE{$j < pop$}
            \STATE Initialize the particle's position $\vec{x}_{j}=\left(s_{\text{mul}}^{(j)}, S_{\text{mul}}^{(j)}\right)$, velocity $\vec{v}_{j}=\left(v_{s}^{(j)},v_{S}^{(j)}\right)$ with uniform distribution.
            \STATE Compute $\left\{y_{it}\right\}_{t=t^*+1}^{T}$ and $\left\{y_{sit}\right\}_{t=t^*+1}^{T}$ (\ref{r3}), (\ref{r4}).
            \IF{$y_{it},y_{sit}>0$ for all $s,i,t$ (no drug shortage)}
            \STATE Initialize the particle's best known position to its initial position: $\vec{p}_{j}=\vec{x}_{j}$
            \IF{$\tilde{f}_{2}\left(\vec{p}_{j}\right)<\tilde{f}_{2}\left(\vec{g}\right)$}
            \STATE Update the swarm's best known position: $\vec{g}=\vec{p}_{j}$
            \ENDIF
            \STATE j=j+1
            \ENDIF
            \ENDWHILE
            \WHILE{$m < i_{\text{max}}$}
            \FOR{$j$ in $1:pop$}
            \STATE Update the particle's velocity: $\vec{v}_{j}=\omega \vec{v}_{j} + c_{1}U(0,1)\left(\vec{p}_{j}-\vec{x}_{j}\right)+c_{2}U(0,1)\left(\vec{g}-\vec{x}_{j}\right)$
            \STATE Update the particle's position: $\vec{x}_{j}=\vec{x}_{j}+\vec{v}_{j}$
            \IF{$\tilde{f}_{2}\left(\vec{x}_{j}\right)<\tilde{f}_{2}\left(\vec{p}_{j}\right)$}
            \STATE Update the particle's best known position: $\vec{p}_{j}=\vec{x}_{j}$
            \IF{$\tilde{f}_{2}\left(\vec{p}_{j}\right)<\tilde{f}_{2}\left(\vec{g}\right)$}
            \STATE Update the swarm's best known position: $\vec{g}=\vec{p}_{j}$
            \ENDIF
            \ENDIF
            \ENDFOR
            \ENDWHILE
		\ENSURE Optimized decision variables $\vec{g}=\left(s_{\text{mul}}^{\text{best}}, S_{\text{mul}}^{\text{best}}\right)$.
	\end{algorithmic}  
\end{algorithm}
In trial planning stage, particle's position and velocity are three-dimensional, corresponding to three decision variables $x_{\text{mul}}, s_{\text{mul}}$, and $ S_{\text{mul}}$. Given with simulated stochastic inputs $\left\{n_{st}\right\}_{t=1}^{T}$, $\left\{\alpha_{st}\right\}_{t=1}^{T}$, $\left\{D_{t}\right\}_{t=1}^{T}$, and $\left\{\gamma_{it}\right\}_{t=1}^{T}$, the number of patients off treatment $\left\{N_{st}\right\}_{t=1}^{T}$, enrollment and supply chain status $\left\{\delta_{t}\right\}_{t=1}^{T}, \left\{\theta_{t}\right\}_{t=1}^{T}$ can be computed sequentially by (\ref{r1}), (\ref{r2}). Drug consumption $\left\{d_{sit}\right\}_{t=1}^{T}$ can be obtained from (\ref{consumptionestimation}). Then, for any particle's position $\vec{x}=\left(x_{\text{mul}}, s_{\text{mul}}, S_{\text{mul}}\right)$, since we can get recommended production amount and resupply thresholds $x_{i}, s_{si}, S_{si}$ by (\ref{feature1.1}), (\ref{feature1.2}), (\ref{feature1.3}), all other parameters in TABLE \ref{taboutput}, $\left\{y_{it}\right\}_{t=1}^{T}, \left\{y_{sit}\right\}_{t=1}^{T}, u_{si}, \left\{u_{sit}\right\}_{t=1}^{T}, v_s, \left\{v_{st}\right\}_{t=1}^{T}$ can be computed sequentially by (\ref{r3}), (\ref{r4}), (\ref{vs}), and (\ref{us}). And we know the value of objective function $f_{1}$ by plugging them in (\ref{obj1}). In summary, in trial planning stage, for any given stochastic inputs $\left\{n_{st}\right\}_{t=1}^{T}$, $\left\{\alpha_{st}\right\}_{t=1}^{T}$, $\left\{D_{t}\right\}_{t=1}^{T}$, and $\left\{\gamma_{it}\right\}_{t=1}^{T}$, a particle's position $\vec{x}=\left(x_{\text{mul}}, s_{\text{mul}}, S_{\text{mul}}\right)$ corresponds uniquely with an objective value. Thus, we can perform particle swarm optimization as Algorithm \ref{alg1}. Note that when initializing the particle’s position, a particle will be accepted only if the corresponding $y_{it},y_{sit}>0$ for all $s,i,t$, which means there is no drug shortage during the trial, since we consider to prevent shutdown as much as possible.\\
In trial monitoring stage, suppose current time point is $t^*$. Particle's position and velocity are now two-dimensional, corresponding to two decision variables $s_{\text{mul}}$, and $ S_{\text{mul}}$, since the production amount has been fixed. Also, the stochastic inputs and supply chain status have been partially observed, which means $\left\{n_{st}\right\}_{t=1}^{t^*}$, $\left\{\alpha_{st}\right\}_{t=1}^{t^*}$, $\left\{D_{t}\right\}_{t=1}^{t^*}$, $\left\{\gamma_{it}\right\}_{t=1}^{t^*}$, $\left\{N_{st}\right\}_{t=1}^{t^*}$, $\left\{\delta_{t}\right\}_{t=1}^{t^*}, \left\{\theta_{t}\right\}_{t=1}^{t^*+\tau}$, $\left\{d_{sit}\right\}_{t=1}^{t^*}$, $\left\{y_{it}\right\}_{t=1}^{t^*}, \left\{y_{sit}\right\}_{t=1}^{t^*}, u_{si}, \left\{u_{sit}\right\}_{t=1}^{t^*}, v_s, \left\{v_{st}\right\}_{t=1}^{t^*}$ are fixed and known parameters. Thus, with partially simulated stochastic inputs $\left\{n_{st}\right\}_{t=t^*+1}^{T}$, $\left\{\alpha_{st}\right\}_{t=t^*+1}^{T}$, $\left\{D_{t}\right\}_{t=t^*+1}^{T}$, $\left\{\gamma_{it}\right\}_{t=t^*+1}^{T}$, a particle's position $\vec{x}=\left(s_{\text{mul}}, S_{\text{mul}}\right)$ corresponds uniquely with a value of objective function $f_{2}$. The optimization can be conducted in Algorithm \ref{alg2}.\\
\begin{algorithm}[htbp]
	\renewcommand{\algorithmicrequire}{\textbf{Input:}}
	\renewcommand{\algorithmicensure}{\textbf{Output:}}
	\caption{Sequential Supply Chain Optimization (Trial Planning Stage)}
	\label{alg3}
	\begin{algorithmic}[1]
            \REQUIRE Number of scenarios simulated $N_{\text{sim}}^{(1)}$; Distributions and parameters to generate stochastic inputs $n_{st}$, $\alpha_{st}$, $D_{t}$, and $\gamma_{it}$; Time point sets $TI$ and $TR$ for interim analysis and resupply; Particle swarm optimization parameters: population $pop$, dimension $D=3$, maximum number of iterations $i_{\text{max}}$, inertia weight $\omega$, cognitive component $c_1$, social component $c_2$.
            \FOR{$k$ in $1:N_{\text{sim}}^{(1)}$}
            \STATE Generate stochastic sequences $\left\{n_{st}^{k}\right\}_{t=1}^{T}$, $\left\{\alpha_{st}^{k}\right\}_{t=1}^{T}$, $\left\{D_{t}^{k}\right\}_{t=1}^{T}$, and $\left\{\gamma_{it}^{k}\right\}_{t=1}^{T}$ for $s\in S$, $i\in I$.
            \STATE Optimize $x_{\text{mul}}^{k}, s_{\text{mul}}^{k}, S_{\text{mul}}^{k}$ with objective function (\ref{obj1}) by Algorithm \ref{alg1}.
            \STATE Obtain $x_i^{k}$, $s_{si}^{k}$, $S_{si}^{k}$ from $x_{\text{mul}}^{k}, s_{\text{mul}}^{k}, S_{\text{mul}}^{k}$ by (\ref{feature1.1}), (\ref{feature1.2}), (\ref{feature1.3}).
            \ENDFOR
		\STATE Decision variable $x_i$ is the $0.99$ quantile of $\left\{x_{i}^{k}\right\}_{k=1}^{N_{\text{sim}}^{(1)}}$; $s_{si}$ and $S_{si}$ are the median of $\left\{s_{si}^k\right\}_{k=1}^{N_{\text{sim}}^{(1)}}$ and $\left\{S_{si}^k\right\}_{k=1}^{N_{\text{sim}}^{(1)}}$.
            \STATE Obtain predicted total cost by computing objective value (\ref{obj1}).
            \STATE Apply $x_i, s_{s i}, S_{s i}$ to the shipment prior to trial and the next time period: $u_{si}=S_{si}$.
		\ENSURE Decision variables $x_i, s_{si}, S_{si}$ and predicted total cost.
	\end{algorithmic}  
\end{algorithm}

\subsubsection{Trial Planning Stage}
Before the trial starts, since the true values of stochastic inputs, $n_{st}$, $\alpha_{st}$, $D_{t}$, and $\gamma_{it}$ remain unknown, we use a large group of simulated paths to cover as many scenarios as possible, which are generated based on previous observations. Take the number of enrolled patients $n_{st}$ as an example. In most cases, the arrival process $n_{st}$ is assumed to follow a Poisson distribution, and the average arrival rate can be estimated by previous sample mean. For the $k$th simulated scenario, suppose we generate a combination of $\left\{n_{st}^{k}\right\}_{t=1}^{T}$, $\left\{\alpha_{st}^{k}\right\}_{t=1}^{T}$, $\left\{D_{t}^{k}\right\}_{t=1}^{T}$, and $\left\{\gamma_{it}^{k}\right\}_{t=1}^{T}$ for $s\in S$, $i\in I$. Then, by Algorithm \ref{alg1}, we can find corresponding optimal $x_{\text{mul}}^{k}, s_{\text{mul}}^{k}, S_{\text{mul}}^{k}$. Recommended production amount and resupply thresholds $x_i^{k}$, $s_{si}^{k}$, $S_{si}^{k}$ can be obtained by (\ref{feature1.1}), (\ref{feature1.2}), (\ref{feature1.3}). Finally, the decision variable $x_i$ is set to be the $99\%$ quantile of all the optimized $x_i^{k}$. In other words, the production amount is enough for $99\%$ of all scenarios and there will be no drug shortage. We let $s_{si}$, $S_{si}$ be the median of all the optimized $s_{si}^{k}$, $S_{si}^{k}$ to avoid over-sufficient or inadequate supply.
\begin{algorithm}[htbp]
	\renewcommand{\algorithmicrequire}{\textbf{Input:}}
	\renewcommand{\algorithmicensure}{\textbf{Output:}}
	\caption{Sequential Supply Chain Optimization (Monitoring Stage)}
	\label{alg4}
	\begin{algorithmic}[1]
            \REQUIRE Number of scenarios simulated $N_{\text{sim}}^{(2)}$; Distributions and parameters to generate stochastic inputs $n_{st}$, $\alpha_{st}$, $D_{t}$, and $\gamma_{it}$; Time point sets $TI$, $TR$, and $TO$ for interim analysis, resupply, and optimization; Particle swarm optimization parameters: population $pop$, dimension $D=2$, maximum number of iterations $i_{\text{max}}$, inertia weight $\omega$, cognitive component $c_1$, social component $c_2$.
            \STATE Set current time point $t^*=1$.
            \STATE Initialize supply chain parameters: $N_{s1}=0$, $\delta_{1}=1$, $\left\{\theta_t\right\}_{t=1}^{1+\tau}=\vec{1}$.
            \REPEAT
            \STATE Update observed stochastic parameters: $\left\{n_{st}^{\text{ob}}\right\}_{t=1}^{t^*}$, $\left\{\alpha_{st}^{\text{ob}}\right\}_{t=1}^{t^*}$, $\left\{D_{t}^{\text{ob}}\right\}_{t=1}^{t^*}$, and $\left\{\gamma_{it}^{\text{ob}}\right\}_{t=1}^{t^*}$ for $s\in S$, $i\in I$.
            \STATE Update observed supply chain parameters: $\left\{N_{st}\right\}_{t=1}^{t^*}$, $\left\{\delta_{t}\right\}_{t=1}^{t^*}$ , $\left\{\theta_{t}\right\}_{t=1}^{t^*+\tau}$, $\left\{y_{it}\right\}_{t=1}^{t^*}$, $\left\{y_{sit}\right\}_{t=1}^{t^*}$, and consumption $\left\{d_{sit}\right\}_{t=1}^{t^*}$
            \STATE (Optional) Update distributions and parameters to generate stochastic inputs $n_{st}$, $\alpha_{st}$, $D_{t}$, and $\gamma_{it}$ (only when there is evidence indicating deviation).
            \IF{$y_{sit^*}<0$ for any $s$ or $i$ (shutdown happens)}
            \STATE Resupply is triggered immediately: $u_{s i t}=S_{s i}-y_{s i t}$, $v_{s t}=\left\lceil\frac{\sum_{i \in I} V u_{s i t}}{Q_{b o x}}\right\rceil$.
            \ELSIF{$t^*\in TR$ (regular resupply checkups)}
            \STATE $u_{s i t}=\mathbb{I}\left(y_{s i t}<s_{s i}\right)\left(S_{s i}-y_{s i t}\right)$, $v_{s t}=\left\lceil\frac{\sum_{i \in I} V u_{s i t}}{Q_{b o x}}\right\rceil$.
            \ENDIF
            \IF{$t^*\in TO$ (time for optimization)}
            \FOR{$k$ in $1:N_{\text{sim}}^{(2)}$}
            \STATE Generate stochastic sequences $\left\{n_{st}^{k}\right\}_{t=t^*+1}^{T}$, $\left\{\alpha_{st}^{k}\right\}_{t=t^*+1}^{T}$, $\left\{D_{t}^{k}\right\}_{t=t^*+1}^{T}$, and $\left\{\gamma_{it}^{k}\right\}_{t=t^*+1}^{T}$ for $s\in S$, $i\in I$.
            \STATE Combine observations as $\left\{n_{st}^k\right\}_{t=1}^{T}=\left\{\left\{n_{st}^{\text{ob}}\right\}_{t=1}^{t^*},\left\{n_{st}^k\right\}_{t=t^*+1}^{T}\right\}$, $\left\{\alpha_{st}^k\right\}_{t=1}^{T}=\left\{\left\{\alpha_{st}^{\text{ob}}\right\}_{t=1}^{t^*},\left\{\alpha_{st}^k\right\}_{t=t^*+1}^{T}\right\}$, $\left\{D_{t}^k\right\}_{t=1}^{T}=\left\{\left\{D_{t}^{\text{ob}}\right\}_{t=1}^{t^*},\left\{D_{t}^k\right\}_{t=t^*+1}^{T}\right\}$, and $\left\{\gamma_{it}^k\right\}_{t=1}^{T}=\left\{\left\{\gamma_{it}^{\text{ob}}\right\}_{t=1}^{t^*},\left\{\gamma_{it}^k\right\}_{t=t^*+1}^{T}\right\}$.
            \STATE Optimize $s_{\text{mul}}^{k}, S_{\text{mul}}^{k}$ with objective function (\ref{obj2}) by Algorithm \ref{alg2}.
            \STATE Obtain $s_{si}^{k}$, $S_{si}^{k}$ from $s_{\text{mul}}^{k}, S_{\text{mul}}^{k}$ by (\ref{feature2.1}), (\ref{feature2.2}).
            \ENDFOR
		\STATE Decision variables $s_{si}$ and $S_{si}$ are the median of $\left\{s_{si}^k\right\}_{k=1}^{N_{\text{sim}}^{(2)}}$ and $\left\{S_{si}^k\right\}_{k=1}^{N_{\text{sim}}^{(2)}}$.
            \STATE Obtain predicted total cost by computing objective value (\ref{obj2}).
            \STATE Apply updated $s_{s i}, S_{s i}$ in the next time period.
            \ENDIF
            \STATE $t^*=t^*+1$.
            \UNTIL $t^*=T$ or $\theta_{t^*}=0$.
	\end{algorithmic}  
\end{algorithm}
\subsubsection{Monitoring Stage}
After trial begins, the production amount is fixed and viewed as a known parameter. In every time point, we keep updating observed stochastic parameters and supply chain parameters for all clinical sites and treatments. Besides, based on the updated observations, we may adjust the way to generate stochastic inputs $n_{st}$, $\alpha_{st}$, $D_{t}$, and $\gamma_{it}$ if there is strong evidence suggesting existence of deviation. For instance, we assume the number of enrolled patients follows a Poisson distribution with mean $\mu$, and when the hypothesis test for the Poisson mean concludes that there is significant evidence to reject the null hypothesis, then the mean $\mu$ needs to be updated.\\
At current time point $t^*$, if shutdown happens at any clinical site, the resupply will be triggered immediately and the resupply amount is determined by current recommended inventory level $S_{si}$.\\
If it is time for regular resupply checkups, we compare the current inventory level at every site with resupply trigger level $s_{si}$. For those needed to be resupplied, the shipment amount is designated to fill the inventory up to the recommended inventory level $S_{si}$.\\
If it is time for optimization, a large group of stochastic sequences $\left\{n_{st}^{k}\right\}_{t=t^*+1}^{T}$, $\left\{\alpha_{st}^{k}\right\}_{t=t^*+1}^{T}$, $\left\{D_{t}^{k}\right\}_{t=t^*+1}^{T}$, and $\left\{\gamma_{it}^{k}\right\}_{t=t^*+1}^{T}$ will be simulated and combined with previous observations. Then, by plugging them in Algorithm \ref{alg2}, corresponding resupply thresholds $s_{\text{mul}}^{k}, S_{\text{mul}}^{k}$ will be optimized. The decision variables are set to be the median of all the $s_{\text{mul}}^{k}, S_{\text{mul}}^{k}$. The updated thresholds are applied in the following resupply campaign till the next optimization time point.

\begin{table}[htbp]
\centering
\caption{This table provides basic drug supply information.}
\label{tabsupply}
\begin{threeparttable}
\begin{tabular}{ll}
\headrow
\textbf{Supply Chain Settings}&\\
Production cost $c_i^p$ (for 3 treatments) &  $0.5, 25.5, 50.5 \$ /$\textit{dose} \\
Recruitment cost $c^{r}_{s}$ (for 5 sites) &  $2000, 2500, 3000, 3500, 4000 \$ /$\textit{week}\\
Shipping cost $c^{S}_{s}$ (for 5 sites) &  $57, 57, 46 ,46, 46 \$ /$\textit{box} \\
Holding cost at distribution center $c^{h}$ & $0.5 \$ /$\textit{dose}$/$\textit{week} \\
Holding cost $c^{h}_{s}$ (for 5 sites) & $0.5 \$ /$\textit{dose}$/$\textit{week} \\
Disposal/recycle cost $c^{W}_{si}$ (for 5 sites) & $25.55, 25.55, 25.55, 31.05, 31.05 \$ /$\textit{dose}\\
Shipment lead time of all dosages and sites $L$ & $1$ \textit{week}\\
Volume of a dose $V$ & $2.56$ \textit{in}$^3$\\
Capacity of a cold shipping box $Q_{\text{box}}$ & $64$ \textit{in}$^3$\\
 Capacity limit of each site $Q_{s}$ & $3000$ \textit{in}$^3$\\
\hline  
\end{tabular}
\end{threeparttable}
\end{table}

\section{Numerical results}

\subsection{Settings \& Data Generation}
We carry out a simulation study to evaluate the performance of the proposed method. A clinical trial with $3$ treatments and $5$ clinical sites is considered. The stochastic parameters are simulated similarly as in \cite{Chen2019}. To be specific, the number of patients enrolled at each clinical site and time point, $n_{st}$, independently follows a Poisson distribution and the mean is set to be $6$. Patient drop-out rate at each site and time point, $\alpha_{st}$, follows a triangular distribution with mean being $1-\sqrt[\tau]{0.7}$. In this case, average drop-out rate at every week is $16\%$ and overall drop-out rate across all clinical trials is around $30\%$. The initial target sample size is $1000$. When interim analysis occurs, the target sample size $D_{t}$ will increase by a percentage, which is randomly generated from a uniform distribution between $[0,0.05]$. As for the drug consumption, every patient will consume $2, 0.9, 1.1$ doses for treatment $i=1,2,3$ on average, where the drug consumption $\gamma_{it}$ follows the uniform distribution restricted by the condition $\sum_{i\in I}\gamma_{it}=4$. Also, note that the average drug consumption can only change at interim analysis timepoints. In the simulation study, we generate the true value of the stochastic parameters based on the distributions and settings described above. During the trial, only the stochastic parameters prior to current time point can be revealed and viewed as known information in the optimization, while the rest will remain unknown. TABLE \ref{tabsupply} summarizes basic supply chain settings in this simulation study. Plus, time to finish the treatment $\tau$ is set to be $2$ weeks. Penalty parameter on the occurrence of drug shortage $P_{\text{shortage}}=500$.\\
Maximum trial duration is $5$ years. During the trial, there are three important pre-specified time point sets $TI, TR$ and $TO$ for interim analysis, resupply, and optimization. In this simulation study, interim analysis happens every four weeks, starting from week $4$. Regular resupply checkups and optimizations are carried out every four weeks as well. As for the particle swarm optimization algorithm parameters, the number of particles (population) is $20$; the inertia weight $\omega$, the cognitive component $c_1$, the social component $c_2$ are set to be $0.9$, $1.6$, and $1.8$ respectively.

\begin{table}[htbp]
\caption{This table provides recommended production amount  and resupply thresholds in trial planning stage.}
\label{tabplan}
\begin{center} 
\begin{tabular}{ c c c c }
\hline
  & Treatment 1 
  & Treatment 2 
  & Treatment 3 \\
\hline
  & \multicolumn{3}{c}{Recommended production amount (dose)}\\
  & $19887$ & $10013$ & $9928$\\
\hline 
& \multicolumn{3}{c}{Trigger level (dose)}\\
Site 1 & $239$ & $119$ & $121$ \\
Site 2 & $243$ & $122$ & $121$ \\
Site 3 & $247$ & $122$ & $122$ \\
Site 4 & $238$ & $120$ & $122$ \\
Site 5 & $236$ & $118$ & $119$ \\
\hline
& \multicolumn{3}{c}{Recommended inventory level (dose)}\\
Site 1 & $327$ & $163$ & $163$ \\
Site 2 & $329$ & $165$ & $165$ \\
Site 3 & $330$ & $167$ & $165$ \\
Site 4 & $329$ & $165$ & $165$ \\
Site 5 & $328$ & $165$ & $163$ \\
\hline 
Total cost (predicted) & \multicolumn{3}{c}{$2748771\$
$}\\
\hline
\end{tabular}
\end{center}
\end{table}

\subsection{Performance of Sequential Supply Chain Optimization Algorithm}
In this simulation study, the supply chain settings are given and fixed; Time points for interim analysis, resupply, and optimization are pre-specified and known; The stochastic inputs, $n_{st}$, $\alpha_{st}$, $D_{t}$, and $\gamma_{it}$, however, remain unknown in trial planning stage and in the monitoring stage, we only know the true value of the stochastic inputs that are prior to current time point. Priori information about the stochastic inputs are given based on historical data, such as the average number of patients enrolled at each site per week, and will keep being adjusted during the trial.
As shown in TABLE \ref{tabplan}, in trial planning stage, the model provides us with recommended production amount for each treatment, resupply thresholds (including trigger level and recommended inventory level) for each site and treatment.\\
\begin{table}[htbp]
\caption{This table provides current inventory, resupply amount and updated resupply thresholds at week $4$ during trial monitoring stage.}
\label{tabmon1}
\begin{center} 
\begin{tabular}{ c c c c c c c }
\hline
\multicolumn{7}{c}{Check point 2 (week 4)}\\
\hline
& Treatment 1 & Treatment 2 & Treatment 3 & Treatment 1 & Treatment 2 & Treatment 3\\
\hline
& \multicolumn{3}{c}{Current inventory at sites (dose)} & \multicolumn{3}{c}{Updated trigger level (dose)}\\
\hline
Site 1 & $273$ & $125$ & $83$ & $206$ & $102$ & $107$ \\
Site 2 & $299$ & $135$ & $125$ & $197$ & $101$ & $102$ \\
Site 3 & $290$ & $133$ & $115$ & $194$ & $101$ & $107$ \\
Site 4 & $270$ & $111$ & $90$ & $203$ & $102$ & $101$ \\
Site 5 & $265$ & $100$ & $99$ & $199$ & $105$ & $102$ \\
\hline
& \multicolumn{3}{c}{Resupply amount (dose)} & \multicolumn{3}{c}{Updated recommended inventory level (dose)}\\
\hline
Site 1 & $0$ & $0$ & $64$ & $289$ & $143$ & $146$ \\
Site 2 & $0$ & $0$ & $0$ & $279$ & $148$ & $142$ \\
Site 3 & $0$ & $0$ & $0$ & $277$ & $144$ & $142$ \\
Site 4 & $0$ & $0$ & $48$ & $283$ & $142$ & $143$ \\
Site 5 & $0$ & $44$ & $44$ & $280$ & $146$ & $144$ \\
\hline
Total cost& \multicolumn{6}{c}{$2448061\$$}\\
\hline
\end{tabular}
\begin{tablenotes}
\item Zero resupply amount means no resupply needed. 
\end{tablenotes}
\end{center}
\end{table}
In the monitoring stage, since the resupply checkups and optimization happens in the same week (according to the simulation settings), we combine the resupply decisions and optimization results in one table. Note that in algorithm 2, we conduct resupply checkups before optimization. Thus, the resupply amount is determined by the thresholds generated in last optimization. TABLE \ref{tabmon1} shows the resupply amount and updated thresholds at week $4$, which is also the second time point for resupply checkups and optimization. There are $3$ sites needed to be resupplied. In TABLE \ref{tabmon2} (week $52$), all $5$ sites require to be resupplied. Compared with week $4$, we observe greater recommended inventory level in week $52$. This is probably because in later period of trial, the inventory at distribution center drops to a relatively lower level. Hence, the algorithm tends to avoid sufficient supply in case the drug stock at depot may not last until the end of trial. However, the trigger levels in week $52$ are almost the same as week $4$ since any lower threshold may lead to drug shortage at clinical sites. The total cost at TABLE \ref{tabmon1} and \ref{tabmon2} represents predicted total cost after current time point, which will keep decreasing during the trial.\\
\begin{table}[htbp]
\caption{This table provides current inventory, resupply amount and updated resupply thresholds at week $52$.}
\label{tabmon2}
\begin{center} 
\begin{tabular}{ c c c c c c c }
\hline
\multicolumn{7}{c}{Check point 14 (week 52)}\\
\hline
& Treatment 1 & Treatment 2 & Treatment 3 & Treatment 1 & Treatment 2 & Treatment 3\\
\hline
& \multicolumn{3}{c}{Current inventory at sites (dose)} & \multicolumn{3}{c}{Updated trigger level (dose)}\\
\hline
Site 1 & $115$ & $62$ & $53$ & $194$ & $103$ & $106$ \\
Site 2 & $129$ & $49$ & $60$ & $193$ & $97$ & $96$ \\
Site 3 & $186$ & $88$ & $78$ & $194$ & $97$ & $103$ \\
Site 4 & $106$ & $77$ & $32$ & $195$ & $108$ & $98$ \\
Site 5 & $132$ & $66$ & $37$ & $196$ & $103$ & $96$ \\
\hline
& \multicolumn{3}{c}{Resupply amount (dose)} & \multicolumn{3}{c}{Updated recommended inventory level (dose)}\\
\hline
Site 1 & $139$ & $66$ & $83$ & $237$ & $127$ & $131$ \\
Site 2 & $129$ & $70$ & $61$ & $236$ & $118$ & $119$ \\
Site 3 & $69$ & $36$ & $60$ & $227$ & $116$ & $124$ \\
Site 4 & $139$ & $59$ & $94$ & $231$ & $129$ & $122$ \\
Site 5 & $118$ & $66$ & $89$ & $234$ & $126$ & $123$ \\
\hline
Total cost& \multicolumn{6}{c}{$778133.1\$$}\\
\hline
\end{tabular}
\end{center}
\end{table}
\begin{figure}[htbp]
\centering
\includegraphics[width=15cm]{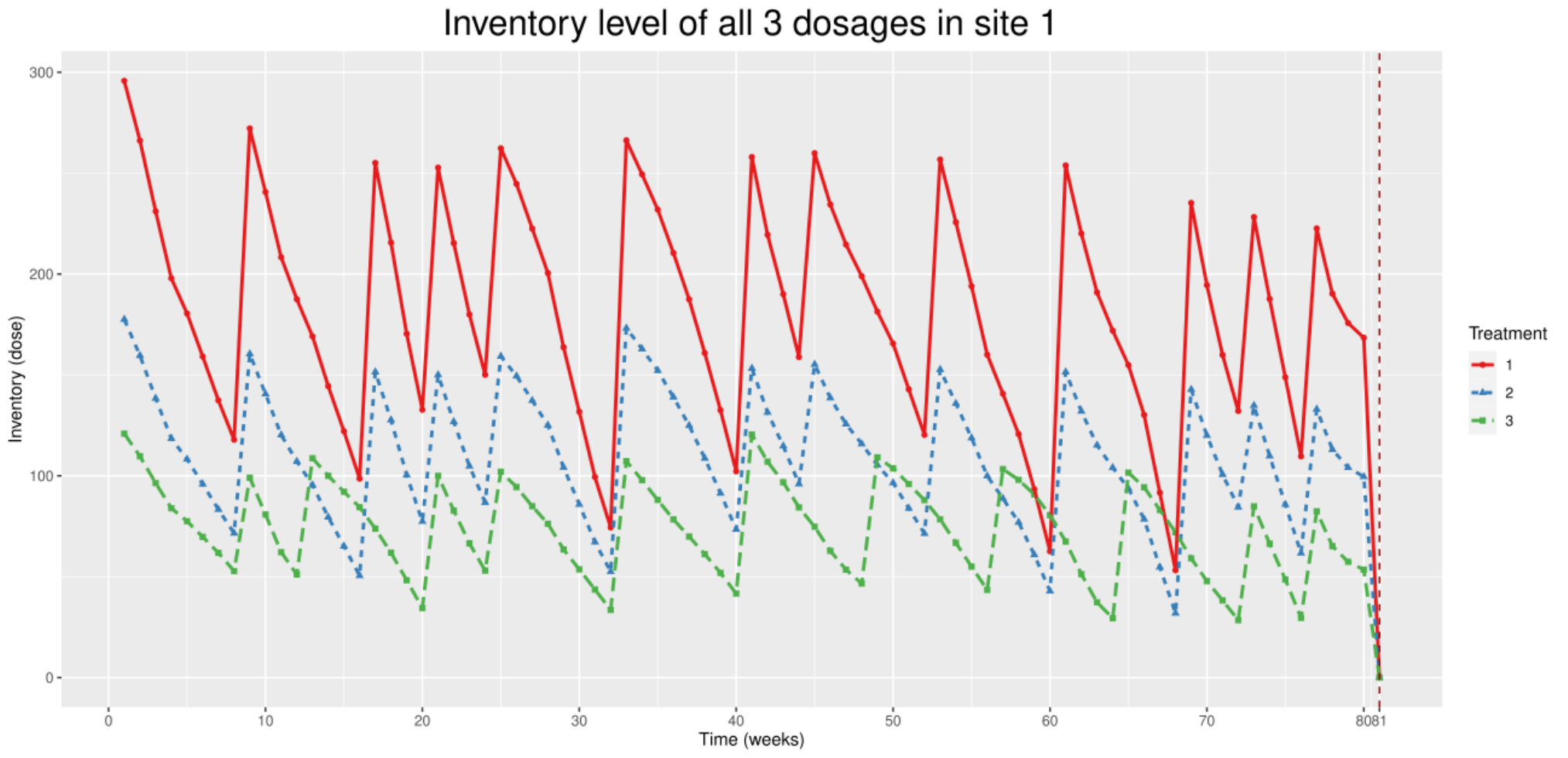}
\caption{This figure shows the changes of inventory level at site $1$. Three  kinds of dosages are indicated by  different colors. Since the trial ends at week $81$, the remaining drugs are recycled and the inventory of all $3$ dosages is zeroed out at week $81$.}
\label{simresult}
\end{figure}
FIGURE \ref{simresult} gives us a visual impression of the changes of the inventory level at clinical sites. In this simulation study, we carry out the regular resupply checkups every four weeks. However, the FIGURE \ref{simresult} shows that no resupply event happened at week $4$. This is because the inventory levels of treatment $1$ $2$, and $3$ are all greater than the corresponding resupply trigger levels at week $4$. The trial ends at week $81$, which implies that the number of patients off treatment $\sum_{s \in S} N_{s, t}$ exceeds the target sample size $D_{t}$ at week $79$. The enrollment is closed after week $79$, but since the patients enrolled at week $79$ require $\tau=2$ more weeks to finish the treatment. The supply chain remains open until week $81$. After that, we recycle all the drugs at clinical sites and the inventory level drops to zero.\\
\begin{table}[htbp]
\centering
\caption{This table provides results in $4$ independent simulation studies.}
\label{tabrepeat}
\begin{threeparttable}
\begin{tabular}{ccccccc}
\headrow
\textbf{Simulation}& Production amount & Drug consumption & Ratio & Shutdown & Total cost$(\$)$ & Duration(week)\\
$1$ & $42660$ & $27039$ & $0.634$ & $0$ & $3967338$ & $95$\\
$2$ & $35869$ & $34537$ & $0.963$ & $0$ & $3767338$ & $107$\\
$3$ & $36194$ & $26749$ & $0.739$ & $0$ & $3171487$ & $80$\\
$4$ & $33131$ & $32196$ & $0.972$ & $0$ & $3481992$ & $102$\\
\hline  
\end{tabular}
\begin{tablenotes}
\item The ratio is computed by the drug consumption dividing the production amount. Shutdown represents the number of shutdown happened during the trial.
\end{tablenotes}
\end{threeparttable}
\end{table}
With the same way to generate stochastic parameters and model settings, we repeat the simulation $4$ times to check the stability of the algorithm. The results are summarized in TABLE \ref{tabrepeat}. In trial planning stage, the Algorithm \ref{alg3} optimizes the recommended production amount for all $3$ dosages, ranged from $33131$ to $42660$. The ratio of drug consumption to production amount reflects the usage rate of the drug produced. As shown in TABLE \ref{tabrepeat}, more than $60\%$ of produced drugs are consumed during the trial, even in the worst scenario, which indicates the algorithm helps to reduce the waste as much as possible. On the other hand, there is no shutdown or drug shortage happened in all the $4$ simulation studies.

\begin{table}[htbp]
\centering
\caption{This table provides results in $4$ independent simulation studies.}
\label{tabsens}
\begin{threeparttable}
\begin{tabular}{ccccccc}
\headrow
True $\mu$ & Production amount & Drug consumption & Ratio & Shutdown & Total cost$(\$)$ & Duration(week)\\
$4$ & $34417$ & $25509$ & $0.741$ & $0$ & $4006271$ & $113$\\
$5$ & $36245$ & $29234$ & $0.807$ & $0$ & $3910958$ & $108$\\
$7$ & $34694$ & $24695$ & $0.712$ & $1$ & $2657960$ & $63$\\
$8$ & $35113$ & $25274$ & $0.720$ & $2$ & $2502882$ & $57$\\
\hline  
\end{tabular}
\begin{tablenotes}
\item The true $\mu$ represents the true value of average enrollment rate, which equals to $4,5,7,8$ patients per week and differs from the assumed Poisson mean $6$ in the optimization algorithm.
\end{tablenotes}
\end{threeparttable}
\end{table}
\subsection{Sensitivity Analysis}
We investigate the robustness property of our model when the enrollment rate is unexpectedly high or low. To be specific, in trial planning stage, the stochastic input $\left\{n_{st}\right\}_{t=1}^{T}$ is simulated from a Poisson distribution with mean $\mu=6$. Then, the Algorithm \ref{alg3} optimizes the production amount $x_i$ and resupply thresholds $s_{si}$ and $ S_{si}$ based on the generated data. However, after trial begins, the true value of $\left\{n_{st}\right\}_{t=1}^{T}$ is actually generated from a Poisson distribution with mean $\mu=4,5,7,8$. We then check if there will be any drug shortage till the end of the trial. Note that in Algorithm \ref{alg4}, the parameters to generate stochastic inputs can be updated based on current observations. In order to test the sensitivity of the sequential supply chain optimization algorithm in both trial planning and monitoring stage, however, we keep the parameters to simulate stochastic inputs being misspecified in both Algorithm \ref{alg3} \& \ref{alg4}. That is, the algorithm will optimize the production amount and resupply thresholds based on the belief that on average, $6$ patients will be enrolled every week, while the true value of the average enrollment rate $\mu=4,5,7,8$ patients per week, respectively.\\
In TABLE \ref{tabsens}, the production amounts are almost the same since in trial planning stage, it is optimized based on the same misspecified Poisson mean of $n_{st}$, which is $6$. And the usage rates of the drug are very close to each other as well. However, with the increasement of the true value of average enrollment rate, we can see that the total cost of the trial keeps dropping. This is because when the enrollment rate becomes unexpectedly high, the trial will end earlier with smaller target sample size. In this case, there will be lower recruitment and holding cost. Meanwhile, drug shortage and trial shutdown may happen since the resupply thresholds are not prepared for such a high enrollment rate. Fortunately, this drug shortage can actually be prevented when implementing the model in a real case. Note that in trial monitoring stage (Algorithm \ref{alg4}), the parameters to simulate stochastic inputs can keep being adjusted based on updated observations.\\
\begin{figure}[htbp]
\centering
\includegraphics[width=15cm]{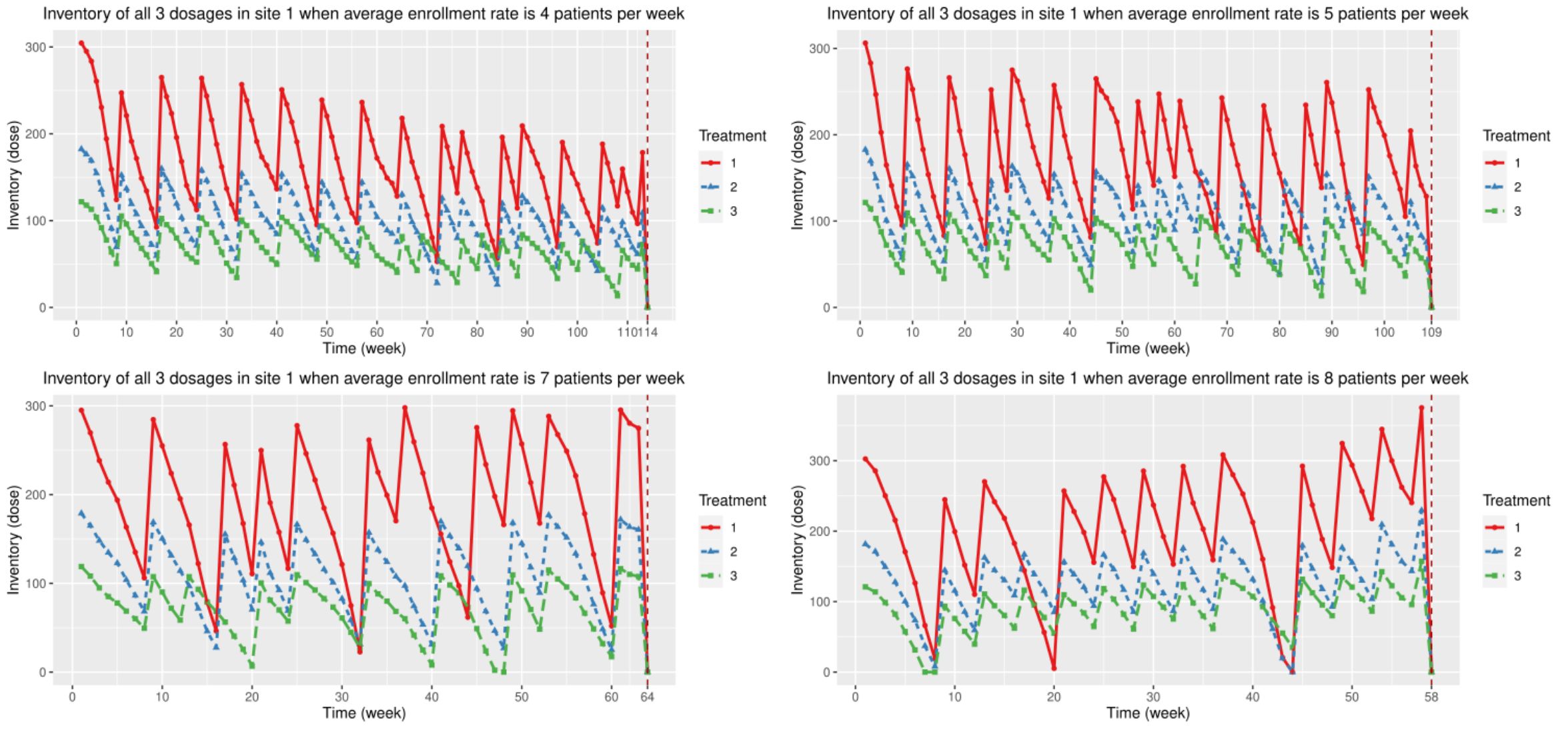}
\caption{This combination of figures compare the changes of inventory level at site $1$ for different misspecified mean enrollment rate. Three kinds of dosages are indicated by  different colors.}
\label{combin}
\end{figure}
FIGURE \ref{combin} provides the comparison of the inventory level at site $1$ among $4$ cases with different enrollment rates. Drug shortage happened once at week $48$ when average enrollment rate is $7$ patients per week, while the shortage happened twice at week $7$ and $44$ when average enrollment rate is $8$ patients per week.

\section{Discussion}
In this paper, we propose a sequential optimization model for the supply chain in both trial planning stage and monitoring stage. With proper assumptions and conditions, the model can predict the production amount needed in the future. Also, resupply thresholds can keep being adjusted during the trial based on updated observations, which means it is an adaptive model for different scenarios. Numerical results show that the model can balance between the supply guarantee and total cost minimization. Also, sensitivity analysis verifies the robustness of our model, where the model works fine even with unexpected patient enrollment rate.\\
Initial value of the particles in PSO algorithm has a great impact on the results. Sometimes the algorithm will not converge because of “Initialization failure”, which means the generated initial value is not located in search space. When that happens, we need to try other ways to identify the search space and find appropriate initial values.\\
We need to be careful when changing the model settings. For example, if the number of weeks apart between resupplies increases from $4$ weeks to $8$ weeks. That is, resupply will happen every $8$ weeks. Then, the capacity limit of site, $Q_s$, probably needs to be increased simultaneously. This is because the optimization algorithm will automatically increase resupply thresholds (resupply trigger level and recommended inventory level) to deal with larger “resupply gap”, which means that the shipment amount will be greater than before. Thus, it is highly likely that the inventory level at sites will exceed the capacity limit of site. In this case, the algorithm will be stuck in the loop.\\
It is of interest to extend the current work in a few directions. Sometimes, when we find the produced drugs are not enough to complete the trial, or if the shelf life of the drug is too short to be produced in one production campaign before trial starts, a second production amount needs to be optimized and determined.

\section*{acknowledgements}
I am profoundly grateful for the continuous support and valuable insights from my managers, Hong Yan, and Zoe Hua. Thanks also to Joseph Ruby for sharing his experience and advice about trial supply chain management.

\printendnotes

\bibliography{sample}

\end{document}